\documentclass[prl,twocolumn,preprintnumbers,amsmath,amssymb]{revtex4}
\usepackage[pdftex,letterpaper, breaklinks=true, pdfborder={0 0 0}, colorlinks=true, linkcolor=blue, pagecolor=blue,citecolor=red, setpagesize=false]{hyperref}
\usepackage{color}
\usepackage{epsfig}
\usepackage{graphicx}
\usepackage{float}
\usepackage{subfigure}
\usepackage{dcolumn}
\usepackage{multirow}

\usepackage{verbatim}
\usepackage{float}
\def\be{\begin{equation}}
\def\ee{\end{equation}}
\def\bea{\begin{eqnarray}}
\def\eea{\end{eqnarray}}
\def\bcb{\begin{color}{blue}}
\def\bcr{\begin{color}{red}}
\def\bcg{\begin{color}{green}}
\def\ec{\end{color}}
\def\bi{\begin{itemize}}
\def\ei{\end{itemize}}
\def\bvb{\begin{verbatim}}
\def\evb{\end{verbatim}}
\newcommand{\unit}[1]{\ensuremath{\, \mathrm{#1}}}

\newcommand\TT{\rule{0pt}{3.0ex}}

\newcommand\BB{\rule[-1.6ex]{0pt}{0pt}}

\begin{document}
\title{False Prediction of Fundamental Properties of Metals by Hybrid Functionals}

\author{Weiwei Gao$^1$}
\author{Tesfaye A. Abtew$^1$}
\author{Tianyi Cai$^1$}
\author{Y. Y. Sun$^2$}
\author{S. B. Zhang$^2$}
\author{Peihong Zhang$^1$}

\affiliation{$^1$Department of Physics, University at Buffalo, State University of New York, Buffalo, New York 14260, USA \\
$^2$Department of Physics, Applied Physics and Astronomy, Rensselaer Polytechnic Institute, Troy, New York 12180, USA}

\date{\today}

\begin{abstract}
The repercussions of an inaccurate account of electronic states near the Fermi 
level $E_F$ by hybrid functionals in predicting several important metallic properties 
are investigated. The difficulties include a vanishing or severely suppressed density of states (DOS) at 
$E_F$, significantly widened valence bandwidth, greatly enhanced electron-phonon (el-ph) deformation potentials, 
and an overestimate of magnetic moment in transition metals.
The erroneously enhanced el-ph coupling calculated by hybrid functionals may lead to a false             
prediction of lattice instability. The main culprit of the problem comes from the simplistic treatment of the 
exchange functional rooted in the original Fock exchange energy. 
The use of a short-ranged Coulomb interaction alleviates
some of the drawbacks but the fundamental issues remain unchanged.
\end{abstract}

\maketitle

As one of the most successful and extensively implemented theories in electronic structure methods, 
the importance of density functional theory (DFT)~\cite{Hohenberg1964,Kohn1965}
can not be overstated.
% ~\cite{Jones1989,Perdew1996,Yin1982,cohen81a,cohen85,yin82b,froyen84,Cramer2009}.
However, despite its tremendous triumph in describing many ground state properties, DFT within
the local density approximation (LDA)~\cite{Kohn1965,Ceperley1980} or the 
generalized gradient approximation (GGA)~\cite{Langreth1983,Becke1988,Perdew1992} 
has serious limitations when it comes to describing excited state properties of materials, a problem that is
often loosely referred to as the LDA band gap problem. 

Strictly speaking, the Kohn-Sham (KS)
eigenvalues cannot be interpreted as the quasiparticle energies in solids. Nevertheless,
there has been much effort put into the development of better energy functionals within DFT that can also
provide a more faithful description of the quasiparticle energies. One class of such functionals,
known as hybrid functionals~\cite{Becke1993}, in which the conventional LDA or GGA exchange-correlation
functional is mixed with a fraction of the Hartree-Fock (HF) exchange energy, has the promises of
providing a better description of a wide range of materials properties
~\cite{Ernzerhof1999,Adamo1999,Muscat2001,Heyd2005,paier06,Dasilva2007,Betzinger2010,Iori2012},
including the band gap of semiconductors. 
There are notable exceptions, however.
For example, the experimental band gap of FeS$_2$, an earth abundant
material with a very high optical absorption coefficient in the visible region, is about
$0.95\unit{eV}$~\cite{Riley2009, Ennaoui1993},
but the Heyd-Scuseria-Ernzerhof (HSE)~\cite{Heyd2003} hybrid functional predicts a band gap
of about $2.6\unit{eV}$.~\cite{Sun2011}

We show that the application of hybrid functionals to metallic systems gives
rise to a range of false or inaccurate results. These include, but not limited to,
1) enlarged valence bandwidth, 
2) vanishing or greatly suppressed density of states (DOS) at the Fermi level,
3) significantly overestimated el-ph deformation potentials, 
4) false predictions of lattice instability for systems (e.g., MgB$_2$) with a strong el-ph coupling, and 
5) enhanced magnetism in transition metals.
The electronic structure near the Fermi level is one of the most important properties of metals.
Many fundamental response, thermal, and transport properties of metals are controlled or strongly affected 
by the electronic structure near the Fermi level.
The inaccurate prediction of the strength of the el-ph coupling, lattice stability, 
and magnetism are just a few manifestations of
the fundamental difficulties that are inherent to hybrid functionals when they are applied to metallic systems.
We believe that the aforementioned problems are serious enough to justify a thorough 
evaluation of the applicability of hybrid functionals to metals and degenerate semiconductors.

Our density functional calculations are performed using the \texttt{Quantum Espresso}~\cite{Giannozzi2009} package.
The GGA functional of Perdew, Burke, and Ernzerhof (PBE)~\cite{Perdew1996} and 
hybrid functionals, both the PBE0~\cite{Perdew1996-PBE0} and the HSE~\cite{Heyd2003} 
functionals, are used in this work. 
Electron-ion interactions are described by norm-conserving pseudopotentials~\cite{Troullier1991}.
The plane wave basis has an energy cutoff of $20\unit{Ry}$ for the calculations of Na,
$50\unit{Ry}$ for MgB$_2$, and $180\unit{Ry}$ for transition metals. For transition metals, all
semicore subshells (i.e., 3$s$, 3$p$, and 3$d$) are included as valence electrons.

\begin{figure}[t!]
\centering
\includegraphics[width=0.45\textwidth]{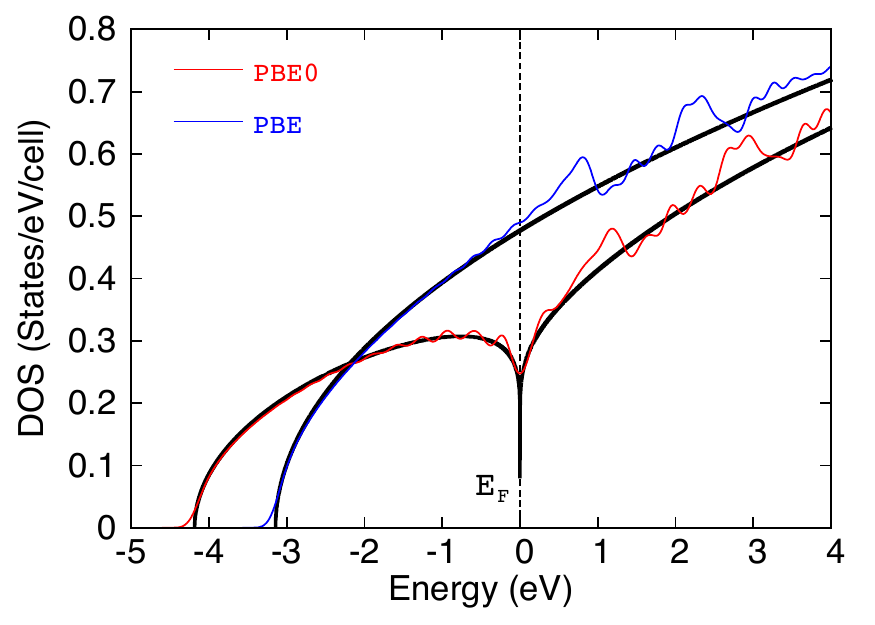}
\caption{(color online) Calculated DOS of sodium using the PBE (in blue) and PBE0 (red) functionals. 
The PBE result agrees well with that of the free-electron model (solid black curve),
and the PBE0 result can be well understood within the HF theory of free electrons (solid
black curve) that shows vanishing DOS at the $E_F$ (shown with the dashed line).}
\label{figure:na-dos}
\end{figure}

We first investigate the electronic structure of metals calculated with
hybrid functional PBE0 using Na as an example.
Figure~\ref{figure:na-dos} compares the DOS of Na calculated
using both the PBE (blue) and the PBE0 (red) functionals.
The DOS are calculated using a very high $k$-point density of $40\times 40\times 40$ 
with a small gaussian smearing parameter of $0.01\unit{eV}$.
It is well-established that the PBE functional gives a satisfactory description of the electronic
structure for a wide range of materials including metals. In fact, the calculated bandwidth ($3.15\unit{eV}$) of Na 
agrees very well with that obtained from the free-electron model (shown in Fig.~\ref{figure:na-dos} with the
black curve superimposed on the PBE result). The PBE (and free-electron model) result also
compares reasonably well with experiment~\cite{Jensen1985,Lyo1988} other than
a well-understood quasiparticle renormalization factor.~\cite{Northrup1989} 
The PBE0 functional, in contrast, gives an overly-exaggerated occupied bandwidth of $4.2\unit{eV}$.
Perhaps a more alarming issue is that the PBE0 functional predicts a vanishing DOS at $E_F$ as shown in
Fig.~\ref{figure:na-dos}.
Note that the singularity is not clearly shown in the figure because of a finite smearing parameter 
used in our calculation. Nevertheless, the significant dip in the DOS at the $E_F$ can be clearly seen.
The abnormal DOS calculated with the PBE0 functional is a direct result
of the HF exchange energy incorporated in the functional.
A simple free-electron calculation using the HF theory 
(with the HF exchange-energy rescaled by a factor of 0.25 as it is used in the PBE0 functional) gives 
nearly identical results as shown with the black curve superimposed on the red curve (PBE0 results).

\begin{figure}[t!]
\centering
\includegraphics[width=0.45\textwidth]{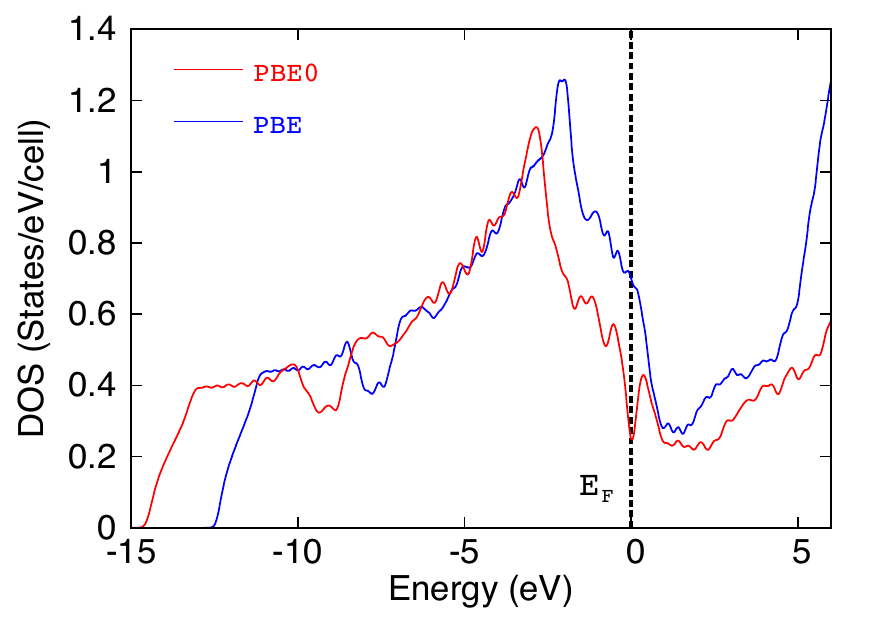}
 \caption{(color online) Density of states of MgB$_2$ calculated using the PBE (blue) and PBE0 (red) functionals.
The dashed line indicates the Fermi level $E_F$.}
  \label{figure:mgb2-dos}
\end{figure}

The Fermi-surface property is perhaps one single most important property of
metals. A wrong description of the Fermi-surface properties, including the DOS and the electron 
energy dispersion near the Fermi level, will have serious consequences
when it comes to predicting other fundamental properties of metals.
In order to demonstrate the repercussion of the incorrect description of the electronic structure
of metals near the Fermi level, we investigate some electron-phonon coupling related 
aspects of metals using MgB$_2$, a well-established multi-gap phonon-mediated 
superconductor~\cite{Nagamatsu2001,Budko2001,An2001, Liu2001,Choi2002}, as an example.
Quantitative predictions of the superconductivity in MgB$_2$, including the precise calculation of
the superconducting transition temperature $T_C$, is one of the celebrated~\cite{An2001, Liu2001,Choi2002}
successes of modern electronic structure methods. This success would not be possible without
an accurate description of the electronic structure near $E_F$.
Figure~\ref{figure:mgb2-dos} shows the DOS of MgB$_2$ calculated using the 
PBE0 (shown in red) and PBE (in blue) functionals.
Similar to the case of Na, the PBE0 functional gives a significantly widened occupied bandwidth
and a greatly suppressed DOS near $E_F$. In fact, a fully-converged PBE0 
calculation (with respect to the $k$-grid density) should give a precisely zero DOS at $E_F$.

\begin{figure}[t!]
\centering
\includegraphics[width=0.35\textwidth]{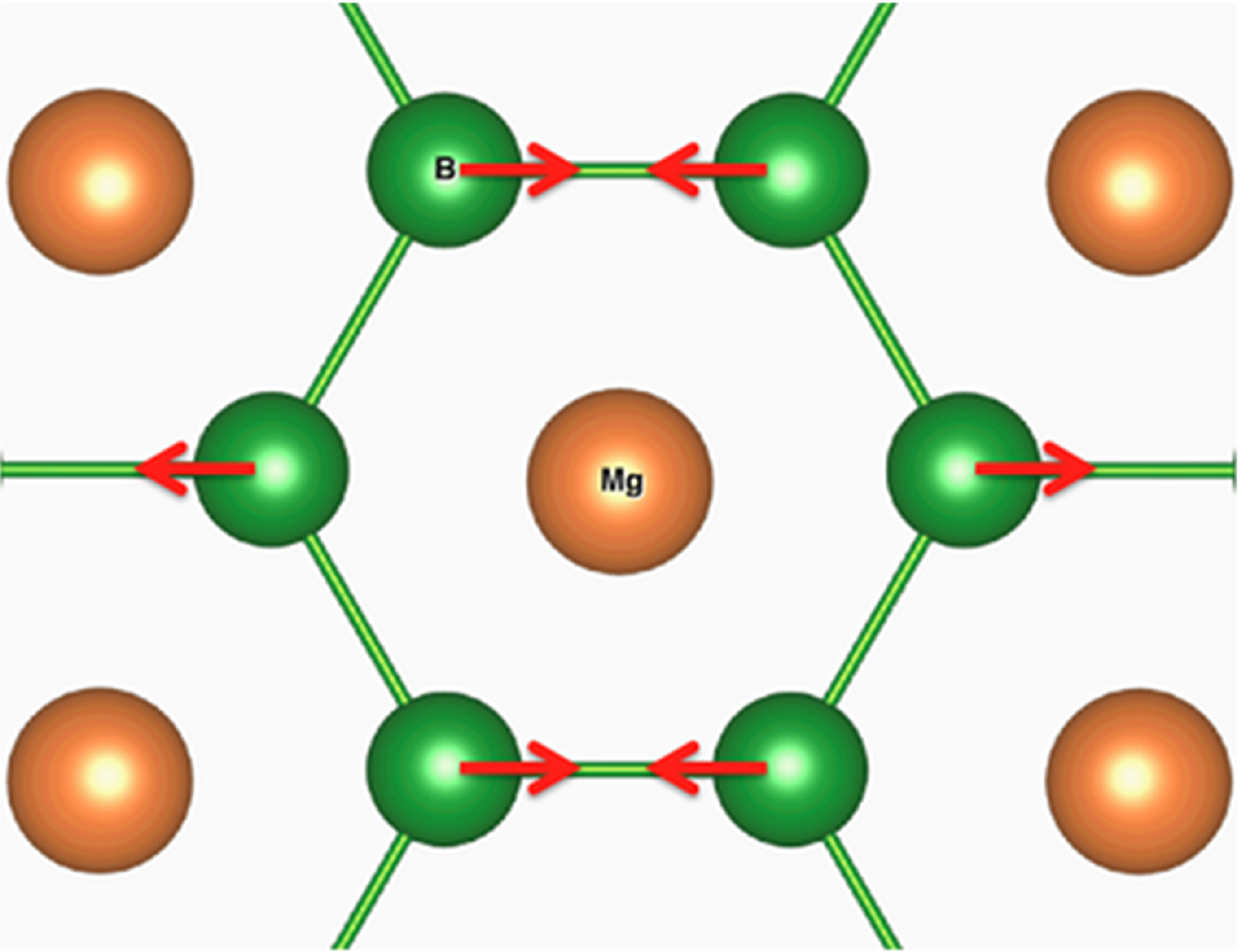}
\caption{Schematics of the polarization vector of one of the $E_{2g}$ phonon modes in MgB$_2$.}
\label{figure:mgb2-E2g-modes}
\end{figure}

%We now examine some of the other serious consequences of the the hybrid functionals when they are applied to
%metal systems by investigating the el-ph deformation potential in MgB$_2$.

\begin{figure}[t!]
\centering
\includegraphics[width=0.45\textwidth]{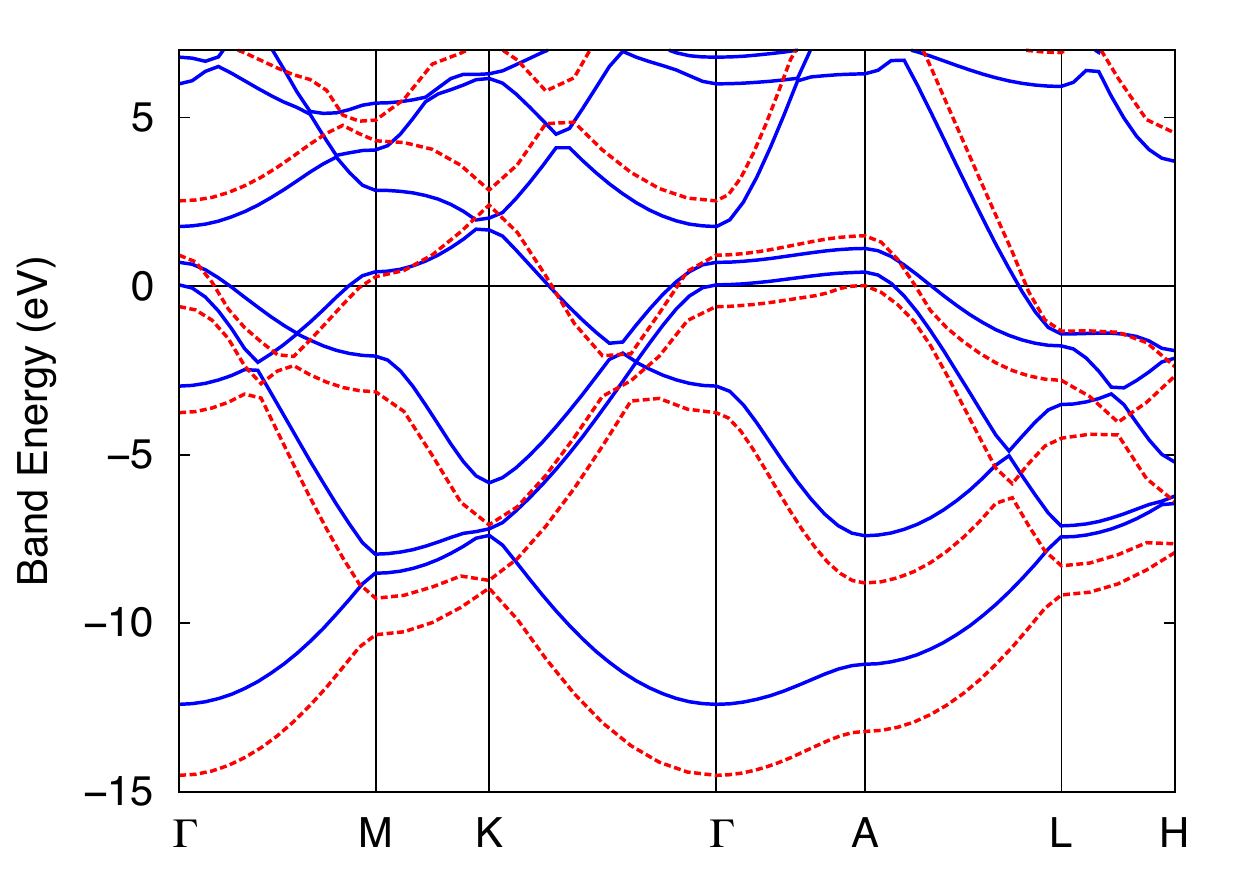}
\caption{Band structures of MgB$_2$ calculated with a frozen-phonon
distortion along one of the E$_{2g}$ modes using the PBE (blue solid line) and PBE0 (red dashed line) functionals.
The magnitude of the distortion is $0.05\unit{a.u.}$}
\label{figure:mgb2-band}
\end{figure}

It is now well-established that the strong coupling between the $E_{2g}$ phonons 
(shown in Fig~\ref{figure:mgb2-E2g-modes}) and 
the boron-derived $p\sigma$ states is largely responsible for the enormously high $T_C$ of MgB$_2$.
Figure~\ref{figure:mgb2-band} shows the band structure of MgB$_2$
calculated with boron atoms displaced by $0.05\unit{a.u.}$ from their equilibrium 
position along one of the $E_{2g}$ phonon modes shown in Fig.~\ref{figure:mgb2-E2g-modes}.
The PBE results are shown with blue solid curves and the PBE0 results are shown with red dashed curves.
The degeneracy of the $p\sigma$ states along the $\Gamma\rightarrow A$ direction is
lifted by the $E_{2g}$ phonon distortion.
An and Pickett~\cite{An2001} used this energy shift (splitting) to calculate the
el-ph deformation potential of the $E_{2g}$ phonons and successfully estimated 
the $T_C$ to be in the range of $32\sim46\unit{K}$ using the PBE functional. 
Our calculated $E_{2g}$ deformation potential $D_{E_{2g}}$ using the PBE functional is about $12.5\unit{eV/\AA}$, which agrees well 
the result of An and Pickett~\cite{An2001}.
The PBE0 functional, on the other hand, predicts a deformation potential of $26.3\unit{eV/\AA}$
which nearly doubles that predicted by the PBE functional. 
This overestimated splitting between the two $p\sigma$ bands with the presence of
the $E_{2g}$ phonon distortion is consistent with the understanding that
hybrid functionals tend to enhance the separation between occupied
and unoccupied electronic states.
Note that the PBE0 band structure is constructed from a self-consistent calculation 
on a $24\times 24\times 24$ uniform $k$-grid.

\begin{figure}[t!]
\centering
\includegraphics[width=0.45\textwidth]{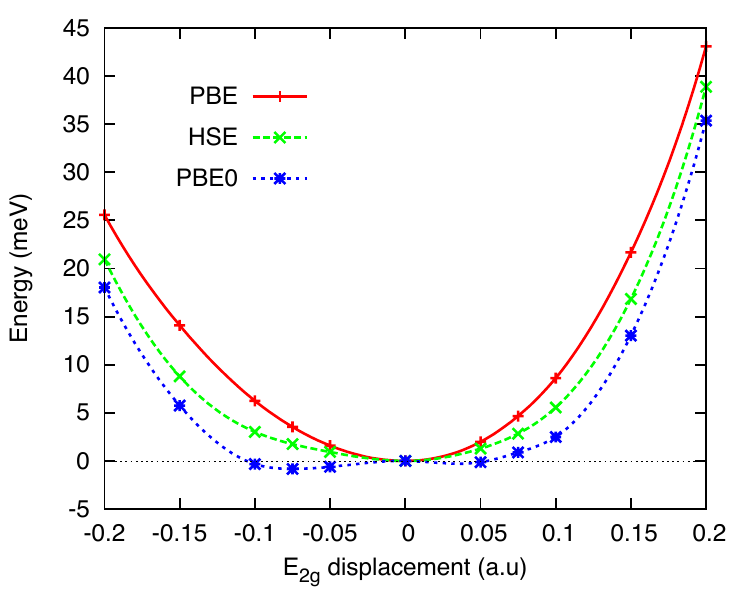}
\caption{Total energy of MgB$_2$ as a function of the magnitude of a frozen-phonon distortion
along one of the E$_{2g}$ phonon modes calculated using the PBE (red), HSE06 (green) and PBE0 (blue)
functional.}
\label{figure:frozenphonon}
\end{figure}

It should be mentioned that the strength of the el-ph coupling in 
MgB$_2$ is so strong that the $E_{2g}$ phonons in MgB$_2$ are greatly renormalized as compared to the
$E_{2g}$ phonons in AlB$_2$.~\cite{Zhang2005}
In fact, the $E_{2g}$ phonons become so soft that the system appears to be close to a 
structural instability. The greatly exaggerated el-ph deformation potential calculated with 
hybrid functionals suggests that these functionals might actually predicts a structural instability for MgB$_2$.
To this end, we have calculated the total energy of MgB$_2$ as a function of the
$E_{2g}$ phonon displacement. 
As shown in Fig.~\ref{figure:frozenphonon}, the potential curve calculated using the PBE functional (shown in red) shows a familiar
third-order anharmonicity.
The potential curve calculated using the HSE functional (green) suggests a greatly
softened $E_{2g}$ phonon compared with the PBE result (red), and the result
from the PBE0 functional shows a double-well structure, indicating an instability
against the $E_{2g}$-like distortions.

The combination of an inaccurate description of the electronic structure
near the Fermi level and the wrong predictions of el-ph coupling and lattice 
stability suggests that hybrid functionals, especially the PBE0 functional, should probably
never be used for metal systems. The inaccurate or false results for simple systems
such as Na or MgB$_2$ predicted by hybrid functionals, however, may not seem to be disturbing enough 
since these functionals are usually {\it not} used for {\it simple} systems.
However, one can also argue that a {\it reliable} energy functional should work equally well for both 
simple and more complex (e.g., strongly correlated) systems. It should be pointed out that
hybrid functionals have been unbiasedly applied to metallic systems,
and the aforementioned issues are general and will be present in
other metallic systems as well.

Finally, we examine further into the limitations and applicability of hybrid functionals to
the understanding of magnetic metals. One of the advantages of
hybrid functionals (compared with the LDA or GGA functionals) is 
that they improve the exchange energy functional and reduces the self-interaction error,
which helps to open the band gap for semiconductors. 
This advantage should be examined with great caution since it also produces
erroneous predictions for a range of metallic properties as discussed above.
Another consequence of including a fraction of the HF exchange energy (either in the
original form such as that in the PBE0 functional or in the screened form as in the case of 
HSE functional) is that it enhances the localization of electronic states and the exchange splitting
between the occupied (or majority spin) and unoccupied (or minority spin) states. 
This enhancement, however, does not guarantee a better description of magnetism.

To illustrate this issue, we have performed spin polarized calculations 
for iron and nickel using the PBE and the HSE hybrid functionals. 
Table~\ref{table:magnetism} compares the calculated magnetic moments of these two systems with experiment.
The HSE06 functional overestimates the spin magnetic moment by about $29\%$ and $61\%$ for Fe and Ni, respectively. 
On the other hand, the magnetism predicted by the PBE functional is in a reasonable agreement with experiment~\cite{crangle1971}.
We mention that similar problem has also been discussed recently~\cite{tran12,jang12} that 
hybrid functionals predict wrong magnetic ground states for Co, Ni, Ph, Pd and Pt.

\begin{table}[t!]
\centering
\caption{Calculated magnetic moments ($\mu_\text{B}$/atom) and relative errors in comparison with experiment~\cite{crangle1971} for Fe and Ni
using PBE and HSE06 functionals.}
\begin{tabular*}{0.48\textwidth}{@{\extracolsep{\fill}}c c c|c c |c c |c}
\hline\hline
\multicolumn{1}{c}\TT{}&
\multicolumn{2}{c|}{PBE}&
%\multicolumn{2}{c|}{PBE0}&
\multicolumn{2}{c|}{HSE06}&
\multicolumn{1}{c}{Experiment}\\
\multicolumn{1}{c}{} &
\multicolumn{1}{c}{M} &
\multicolumn{1}{c|}{Error} &
%\multicolumn{1}{c}{M} &
%\multicolumn{1}{c|}{Error} &
\multicolumn{1}{c}{M} &
\multicolumn{1}{c|}{Error} &
\multicolumn{1}{c}{M} \BB\\
%\multicolumn{1}{c}{} &
%\multicolumn{1}{c}{($\mu_\text{B}$/atom)} &
%\multicolumn{1}{c}{(\%)} &
%\multicolumn{1}{c}{($\mu_\text{B}$/atom)} &
%\multicolumn{1}{c}{(\%)} &
%\multicolumn{1}{c}{($\mu_\text{B}$/atom)} \BB\\

\hline
%		 &PBE		        &PBE0  &HSE06	       		&Experiment         	\\ \hline		 
%\TT Fe		 &2.29 &	2.2\% &- &	-		&2.81	&      25\%	&2.25      \\
%       Ni	 	 &0.70 &	7.7\% &- &	-		&1.00	&	54\%	&0.65	\BB 
\TT Fe		 &2.29 &		3\%   & 2.87	&      29\%	&2.22      \\
  \BB     Ni	 	 &0.70 &		13\% & 1.00	&	61\%	& 0.62 \\

\hline\hline
\end{tabular*}
\label{table:magnetism}
\end{table}

In conclusion, we have shown that there are serious limitations of hybrid functionals 
in predicting fundamental properties of metals.
The inaccurate description of the electron structure of metals,
including a vanishing DOS at the Fermi level, exaggerated
occupied bandwidth, and enhanced splitting between occupied and unoccupied states, 
have profound and adverse effects on the prediction of other important properties of metals.
These difficulties can be traced back to the overly simplified HF-like exchange energy 
included in hybrid functionals which is known to give an the inaccurate description of 
electronic structure near the Fermi level. Introducing a short-ranged
Coulomb potential as in the case of the HSE functional does alleviate
some of the problems, but major difficulties remain unchanged.
Although computationally demanding, hybrid functionals are becoming increasingly popular
with the implicit assumption that they will give a more faithful description 
of the underlying electronic structure. 
Our results raise a serious question regarding the applicability of
hybrid functionals to the study metallic systems.

\section{Acknowledgement}
This work is supported in part by the National Science Foundation under Grant No. DMR-0946404.
Y.S. and S.Z. are supported by the Department of Energy under Grant No. DE-SC0002623.
We acknowledge the computational support provided by the Center for
Computational Research at the University at Buffalo, SUNY.

\bibliographystyle{apsrev}
%\bibliography{MgB2}

\begin{thebibliography}{35}
\expandafter\ifx\csname natexlab\endcsname\relax\def\natexlab#1{#1}\fi
\expandafter\ifx\csname bibnamefont\endcsname\relax
  \def\bibnamefont#1{#1}\fi
\expandafter\ifx\csname bibfnamefont\endcsname\relax
  \def\bibfnamefont#1{#1}\fi
\expandafter\ifx\csname citenamefont\endcsname\relax
  \def\citenamefont#1{#1}\fi
\expandafter\ifx\csname url\endcsname\relax
  \def\url#1{\texttt{#1}}\fi
\expandafter\ifx\csname urlprefix\endcsname\relax\def\urlprefix{URL }\fi
\providecommand{\bibinfo}[2]{#2}
\providecommand{\eprint}[2][]{\url{#2}}

\bibitem[{\citenamefont{Hohenberg and Kohn}(1964)}]{Hohenberg1964}
\bibinfo{author}{\bibfnamefont{P.}~\bibnamefont{Hohenberg}} \bibnamefont{and}
  \bibinfo{author}{\bibfnamefont{W.}~\bibnamefont{Kohn}},
  \bibinfo{journal}{Phys. Rev.} \textbf{\bibinfo{volume}{136}},
  \bibinfo{pages}{B864} (\bibinfo{year}{1964}).

\bibitem[{\citenamefont{Kohn and Sham}(1965)}]{Kohn1965}
\bibinfo{author}{\bibfnamefont{W.}~\bibnamefont{Kohn}} \bibnamefont{and}
  \bibinfo{author}{\bibfnamefont{L.~J.} \bibnamefont{Sham}},
  \bibinfo{journal}{Phys. Rev.} \textbf{\bibinfo{volume}{140}},
  \bibinfo{pages}{A1133} (\bibinfo{year}{1965}).

\bibitem[{\citenamefont{Ceperley and Alder}(1980)}]{Ceperley1980}
\bibinfo{author}{\bibfnamefont{D.~M.} \bibnamefont{Ceperley}} \bibnamefont{and}
  \bibinfo{author}{\bibfnamefont{B.~J.} \bibnamefont{Alder}},
  \bibinfo{journal}{Phys. Rev. Lett.} \textbf{\bibinfo{volume}{45}},
  \bibinfo{pages}{566} (\bibinfo{year}{1980}).

\bibitem[{\citenamefont{Perdew et~al.}(1992)\citenamefont{Perdew, Chevary,
  Vosko, Jackson, Pederson, Singh, and Fiolhais}}]{Perdew1992}
\bibinfo{author}{\bibfnamefont{J.~P.} \bibnamefont{Perdew}},
  \bibinfo{author}{\bibfnamefont{J.~A.} \bibnamefont{Chevary}},
  \bibinfo{author}{\bibfnamefont{S.~H.} \bibnamefont{Vosko}},
  \bibinfo{author}{\bibfnamefont{K.~A.} \bibnamefont{Jackson}},
  \bibinfo{author}{\bibfnamefont{M.~R.} \bibnamefont{Pederson}},
  \bibinfo{author}{\bibfnamefont{D.~J.} \bibnamefont{Singh}}, \bibnamefont{and}
  \bibinfo{author}{\bibfnamefont{C.}~\bibnamefont{Fiolhais}},
  \bibinfo{journal}{Phys. Rev. B} \textbf{\bibinfo{volume}{46}},
  \bibinfo{pages}{6671} (\bibinfo{year}{1992}).

\bibitem[{\citenamefont{Becke}(1988)}]{Becke1988}
\bibinfo{author}{\bibfnamefont{A.~D.} \bibnamefont{Becke}},
  \bibinfo{journal}{Phys. Rev. A.} \textbf{\bibinfo{volume}{38}},
  \bibinfo{pages}{3098} (\bibinfo{year}{1988}).

\bibitem[{\citenamefont{Langreth and Mehl}(1983)}]{Langreth1983}
\bibinfo{author}{\bibfnamefont{D.~C.} \bibnamefont{Langreth}} \bibnamefont{and}
  \bibinfo{author}{\bibfnamefont{M.~J.} \bibnamefont{Mehl}},
  \bibinfo{journal}{Phys. Rev. B} \textbf{\bibinfo{volume}{28}},
  \bibinfo{pages}{1809} (\bibinfo{year}{1983}).

\bibitem[{\citenamefont{Becke}(1993)}]{Becke1993}
\bibinfo{author}{\bibfnamefont{A.~D.} \bibnamefont{Becke}},
  \bibinfo{journal}{J. Chem. Phys.} \textbf{\bibinfo{volume}{98}},
  \bibinfo{pages}{1372Ð} (\bibinfo{year}{1993}).

\bibitem[{\citenamefont{Paier et~al.}(2006)\citenamefont{Paier, Marsman,
  Hummer, Kresse, Gerber, and Angyan}}]{paier06}
\bibinfo{author}{\bibfnamefont{J.}~\bibnamefont{Paier}},
  \bibinfo{author}{\bibfnamefont{M.}~\bibnamefont{Marsman}},
  \bibinfo{author}{\bibfnamefont{K.}~\bibnamefont{Hummer}},
  \bibinfo{author}{\bibfnamefont{G.}~\bibnamefont{Kresse}},
  \bibinfo{author}{\bibfnamefont{I.~C.} \bibnamefont{Gerber}},
  \bibnamefont{and} \bibinfo{author}{\bibfnamefont{J.~G.}
  \bibnamefont{Angyan}}, \bibinfo{journal}{J. Chem. Phys.}
  \textbf{\bibinfo{volume}{124}}, \bibinfo{eid}{154709} (\bibinfo{year}{2006}).

\bibitem[{\citenamefont{Ernzerhof and Scuseria}(1999)}]{Ernzerhof1999}
\bibinfo{author}{\bibfnamefont{M.}~\bibnamefont{Ernzerhof}} \bibnamefont{and}
  \bibinfo{author}{\bibfnamefont{G.~E.} \bibnamefont{Scuseria}},
  \bibinfo{journal}{J. Chem. Phys.} \textbf{\bibinfo{volume}{110}},
  \bibinfo{pages}{5029Ð} (\bibinfo{year}{1999}).

\bibitem[{\citenamefont{Adamo and Barone}(1999)}]{Adamo1999}
\bibinfo{author}{\bibfnamefont{C.}~\bibnamefont{Adamo}} \bibnamefont{and}
  \bibinfo{author}{\bibfnamefont{V.}~\bibnamefont{Barone}},
  \bibinfo{journal}{J. Chem. Phys.} \textbf{\bibinfo{volume}{110}},
  \bibinfo{pages}{6158Ð} (\bibinfo{year}{1999}).

\bibitem[{\citenamefont{Da~Silva et~al.}(2007)\citenamefont{Da~Silva,
  Ganduglia-Pirovano, Sauer, Bayer, and Kresse}}]{Dasilva2007}
\bibinfo{author}{\bibfnamefont{J.~L.~F.} \bibnamefont{Da~Silva}},
  \bibinfo{author}{\bibfnamefont{M.~V.} \bibnamefont{Ganduglia-Pirovano}},
  \bibinfo{author}{\bibfnamefont{J.}~\bibnamefont{Sauer}},
  \bibinfo{author}{\bibfnamefont{V.}~\bibnamefont{Bayer}}, \bibnamefont{and}
  \bibinfo{author}{\bibfnamefont{G.}~\bibnamefont{Kresse}},
  \bibinfo{journal}{Phys. Rev. B} \textbf{\bibinfo{volume}{75}},
  \bibinfo{pages}{045121} (\bibinfo{year}{2007}).

\bibitem[{\citenamefont{Muscat et~al.}(2001)\citenamefont{Muscat, Wander, and
  Harrison}}]{Muscat2001}
\bibinfo{author}{\bibfnamefont{J.}~\bibnamefont{Muscat}},
  \bibinfo{author}{\bibfnamefont{A.}~\bibnamefont{Wander}}, \bibnamefont{and}
  \bibinfo{author}{\bibfnamefont{N.}~\bibnamefont{Harrison}},
  \bibinfo{journal}{Chem. Phys. Lett.} \textbf{\bibinfo{volume}{342}},
  \bibinfo{pages}{397} (\bibinfo{year}{2001}), ISSN \bibinfo{issn}{0009-2614}.

\bibitem[{\citenamefont{Heyd et~al.}(2005)\citenamefont{Heyd, Peralta,
  Scuseria, and Martin}}]{Heyd2005}
\bibinfo{author}{\bibfnamefont{J.}~\bibnamefont{Heyd}},
  \bibinfo{author}{\bibfnamefont{J.~E.} \bibnamefont{Peralta}},
  \bibinfo{author}{\bibfnamefont{G.~E.} \bibnamefont{Scuseria}},
  \bibnamefont{and} \bibinfo{author}{\bibfnamefont{R.~L.}
  \bibnamefont{Martin}}, \bibinfo{journal}{J. Chem. Phys.}
  \textbf{\bibinfo{volume}{123}}, \bibinfo{eid}{174101} (\bibinfo{year}{2005}).

\bibitem[{\citenamefont{Betzinger et~al.}(2010)\citenamefont{Betzinger,
  Friedrich, and Blugel}}]{Betzinger2010}
\bibinfo{author}{\bibfnamefont{M.}~\bibnamefont{Betzinger}},
  \bibinfo{author}{\bibfnamefont{C.}~\bibnamefont{Friedrich}},
  \bibnamefont{and} \bibinfo{author}{\bibfnamefont{S.}~\bibnamefont{Blugel}},
  \bibinfo{journal}{Phys. Rev. B} \textbf{\bibinfo{volume}{81}},
  \bibinfo{pages}{195117} (\bibinfo{year}{2010}).

\bibitem[{\citenamefont{Iori et~al.}(2012)\citenamefont{Iori, Gatti, and
  Rubio}}]{Iori2012}
\bibinfo{author}{\bibfnamefont{F.}~\bibnamefont{Iori}},
  \bibinfo{author}{\bibfnamefont{M.}~\bibnamefont{Gatti}}, \bibnamefont{and}
  \bibinfo{author}{\bibfnamefont{A.}~\bibnamefont{Rubio}},
  \bibinfo{journal}{Phys. Rev. B} \textbf{\bibinfo{volume}{85}},
  \bibinfo{pages}{115129} (\bibinfo{year}{2012}).

\bibitem[{\citenamefont{Murphy and Strongin}(2009)}]{Riley2009}
\bibinfo{author}{\bibfnamefont{R.}~\bibnamefont{Murphy}} \bibnamefont{and}
  \bibinfo{author}{\bibfnamefont{D.~R.} \bibnamefont{Strongin}},
  \bibinfo{journal}{Surf. Sci. Rep.} \textbf{\bibinfo{volume}{64}},
  \bibinfo{pages}{1} (\bibinfo{year}{2009}).

\bibitem[{\citenamefont{Ennaoui et~al.}(1993)\citenamefont{Ennaoui, Fiechter,
  Pettenkofer, Alonso-Vante, Bker, Bronold, Hpfner, and
  Tributsch}}]{Ennaoui1993}
\bibinfo{author}{\bibfnamefont{A.}~\bibnamefont{Ennaoui}},
  \bibinfo{author}{\bibfnamefont{S.}~\bibnamefont{Fiechter}},
  \bibinfo{author}{\bibfnamefont{C.}~\bibnamefont{Pettenkofer}},
  \bibinfo{author}{\bibfnamefont{N.}~\bibnamefont{Alonso-Vante}},
  \bibinfo{author}{\bibfnamefont{K.}~\bibnamefont{Bker}},
  \bibinfo{author}{\bibfnamefont{M.}~\bibnamefont{Bronold}},
  \bibinfo{author}{\bibfnamefont{C.}~\bibnamefont{Hpfner}}, \bibnamefont{and}
  \bibinfo{author}{\bibfnamefont{H.}~\bibnamefont{Tributsch}},
  \bibinfo{journal}{Sol. Energ. Mat. Sol. C} \textbf{\bibinfo{volume}{29}},
  \bibinfo{pages}{289} (\bibinfo{year}{1993}).

\bibitem[{\citenamefont{Heyd et~al.}(2003)\citenamefont{Heyd, Scuseria, and
  Ernzerhof}}]{Heyd2003}
\bibinfo{author}{\bibfnamefont{J.}~\bibnamefont{Heyd}},
  \bibinfo{author}{\bibfnamefont{G.~E.} \bibnamefont{Scuseria}},
  \bibnamefont{and}
  \bibinfo{author}{\bibfnamefont{M.}~\bibnamefont{Ernzerhof}},
  \bibinfo{journal}{J. Chem. Phys.} \textbf{\bibinfo{volume}{118}},
  \bibinfo{pages}{8207} (\bibinfo{year}{2003}).

\bibitem[{\citenamefont{Sun et~al.}(2011)\citenamefont{Sun, Chan, and
  Ceder}}]{Sun2011}
\bibinfo{author}{\bibfnamefont{R.}~\bibnamefont{Sun}},
  \bibinfo{author}{\bibfnamefont{M.~K.~Y.} \bibnamefont{Chan}},
  \bibnamefont{and} \bibinfo{author}{\bibfnamefont{G.}~\bibnamefont{Ceder}},
  \bibinfo{journal}{Phys. Rev. B} \textbf{\bibinfo{volume}{83}},
  \bibinfo{pages}{235311} (\bibinfo{year}{2011}).

\bibitem[{\citenamefont{Giannozzi et~al.}(2009)\citenamefont{Giannozzi, Baroni,
  Bonini, Calandra, Car, Cavazzoni, Ceresoli, Chiarotti, Cococcioni, Dabo
  et~al.}}]{Giannozzi2009}
\bibinfo{author}{\bibfnamefont{P.}~\bibnamefont{Giannozzi}},
  \bibinfo{author}{\bibfnamefont{S.}~\bibnamefont{Baroni}},
  \bibinfo{author}{\bibfnamefont{N.}~\bibnamefont{Bonini}},
  \bibinfo{author}{\bibfnamefont{M.}~\bibnamefont{Calandra}},
  \bibinfo{author}{\bibfnamefont{R.}~\bibnamefont{Car}},
  \bibinfo{author}{\bibfnamefont{C.}~\bibnamefont{Cavazzoni}},
  \bibinfo{author}{\bibfnamefont{D.}~\bibnamefont{Ceresoli}},
  \bibinfo{author}{\bibfnamefont{G.~L.} \bibnamefont{Chiarotti}},
  \bibinfo{author}{\bibfnamefont{M.}~\bibnamefont{Cococcioni}},
  \bibinfo{author}{\bibfnamefont{I.}~\bibnamefont{Dabo}}, \bibnamefont{et~al.},
  \bibinfo{journal}{J. Phys.: Condens. Matter} \textbf{\bibinfo{volume}{21}},
  \bibinfo{pages}{395502} (\bibinfo{year}{2009}).

\bibitem[{\citenamefont{Perdew et~al.}(1996{\natexlab{a}})\citenamefont{Perdew,
  Burke, and Ernzerhof}}]{Perdew1996}
\bibinfo{author}{\bibfnamefont{J.~P.} \bibnamefont{Perdew}},
  \bibinfo{author}{\bibfnamefont{K.}~\bibnamefont{Burke}}, \bibnamefont{and}
  \bibinfo{author}{\bibfnamefont{M.}~\bibnamefont{Ernzerhof}},
  \bibinfo{journal}{Phys. Rev. Lett.} \textbf{\bibinfo{volume}{77}},
  \bibinfo{pages}{3865} (\bibinfo{year}{1996}{\natexlab{a}}).

\bibitem[{\citenamefont{Perdew et~al.}(1996{\natexlab{b}})\citenamefont{Perdew,
  Ernzerhof, and Burke}}]{Perdew1996-PBE0}
\bibinfo{author}{\bibfnamefont{J.~P.} \bibnamefont{Perdew}},
  \bibinfo{author}{\bibfnamefont{M.}~\bibnamefont{Ernzerhof}},
  \bibnamefont{and} \bibinfo{author}{\bibfnamefont{K.}~\bibnamefont{Burke}},
  \bibinfo{journal}{J. Chem. Phys.} \textbf{\bibinfo{volume}{105}},
  \bibinfo{pages}{9982} (\bibinfo{year}{1996}{\natexlab{b}}).

\bibitem[{\citenamefont{Troullier and Martins}(1991)}]{Troullier1991}
\bibinfo{author}{\bibfnamefont{N.}~\bibnamefont{Troullier}} \bibnamefont{and}
  \bibinfo{author}{\bibfnamefont{J.~L.} \bibnamefont{Martins}},
  \bibinfo{journal}{Phys. Rev. B} \textbf{\bibinfo{volume}{43}},
  \bibinfo{pages}{1993} (\bibinfo{year}{1991}).

\bibitem[{\citenamefont{Jensen and Plummer}(1985)}]{Jensen1985}
\bibinfo{author}{\bibfnamefont{E.}~\bibnamefont{Jensen}} \bibnamefont{and}
  \bibinfo{author}{\bibfnamefont{E.~W.} \bibnamefont{Plummer}},
  \bibinfo{journal}{Phys. Rev. Lett.} \textbf{\bibinfo{volume}{55}},
  \bibinfo{pages}{1912} (\bibinfo{year}{1985}).

\bibitem[{\citenamefont{Lyo and Plummer}(1988)}]{Lyo1988}
\bibinfo{author}{\bibfnamefont{I.-W.} \bibnamefont{Lyo}} \bibnamefont{and}
  \bibinfo{author}{\bibfnamefont{E.~W.} \bibnamefont{Plummer}},
  \bibinfo{journal}{Phys. Rev. Lett.} \textbf{\bibinfo{volume}{60}},
  \bibinfo{pages}{1558} (\bibinfo{year}{1988}).

\bibitem[{\citenamefont{Northrup et~al.}(1989)\citenamefont{Northrup,
  Hybertsen, and Louie}}]{Northrup1989}
\bibinfo{author}{\bibfnamefont{J.~E.} \bibnamefont{Northrup}},
  \bibinfo{author}{\bibfnamefont{M.~S.} \bibnamefont{Hybertsen}},
  \bibnamefont{and} \bibinfo{author}{\bibfnamefont{S.~G.} \bibnamefont{Louie}},
  \bibinfo{journal}{Phys. Rev. B} \textbf{\bibinfo{volume}{39}},
  \bibinfo{pages}{8198} (\bibinfo{year}{1989}).

\bibitem[{\citenamefont{Nagamatsu et~al.}(2001)\citenamefont{Nagamatsu,
  Nakagawa, Muranaka, Zenitani, and Akimitsu}}]{Nagamatsu2001}
\bibinfo{author}{\bibfnamefont{J.}~\bibnamefont{Nagamatsu}},
  \bibinfo{author}{\bibfnamefont{N.}~\bibnamefont{Nakagawa}},
  \bibinfo{author}{\bibfnamefont{T.}~\bibnamefont{Muranaka}},
  \bibinfo{author}{\bibfnamefont{Y.}~\bibnamefont{Zenitani}}, \bibnamefont{and}
  \bibinfo{author}{\bibfnamefont{J.}~\bibnamefont{Akimitsu}},
  \bibinfo{journal}{Nature} \textbf{\bibinfo{volume}{410}}, \bibinfo{pages}{63}
  (\bibinfo{year}{2001}).

\bibitem[{\citenamefont{Bud'ko et~al.}(2001)\citenamefont{Bud'ko, Lapertot,
  Petrovic, Cunningham, Anderson, and Canfield}}]{Budko2001}
\bibinfo{author}{\bibfnamefont{S.~L.} \bibnamefont{Bud'ko}},
  \bibinfo{author}{\bibfnamefont{G.}~\bibnamefont{Lapertot}},
  \bibinfo{author}{\bibfnamefont{C.}~\bibnamefont{Petrovic}},
  \bibinfo{author}{\bibfnamefont{C.~E.} \bibnamefont{Cunningham}},
  \bibinfo{author}{\bibfnamefont{N.}~\bibnamefont{Anderson}}, \bibnamefont{and}
  \bibinfo{author}{\bibfnamefont{P.~C.} \bibnamefont{Canfield}},
  \bibinfo{journal}{Phys. Rev. Lett.} \textbf{\bibinfo{volume}{86}},
  \bibinfo{pages}{1877} (\bibinfo{year}{2001}).

\bibitem[{\citenamefont{An and Pickett}(2001)}]{An2001}
\bibinfo{author}{\bibfnamefont{J.~M.} \bibnamefont{An}} \bibnamefont{and}
  \bibinfo{author}{\bibfnamefont{W.~E.} \bibnamefont{Pickett}},
  \bibinfo{journal}{Phys. Rev. Lett.} \textbf{\bibinfo{volume}{86}},
  \bibinfo{pages}{4366} (\bibinfo{year}{2001}).

\bibitem[{\citenamefont{Liu et~al.}(2001)\citenamefont{Liu, Mazin, and
  Kortus}}]{Liu2001}
\bibinfo{author}{\bibfnamefont{A.~Y.} \bibnamefont{Liu}},
  \bibinfo{author}{\bibfnamefont{I.~I.} \bibnamefont{Mazin}}, \bibnamefont{and}
  \bibinfo{author}{\bibfnamefont{J.}~\bibnamefont{Kortus}},
  \bibinfo{journal}{Phys. Rev. Lett.} \textbf{\bibinfo{volume}{87}},
  \bibinfo{pages}{087005} (\bibinfo{year}{2001}).

\bibitem[{\citenamefont{Choi et~al.}(2002)\citenamefont{Choi, Roundy, Sun,
  Cohen, and Louie}}]{Choi2002}
\bibinfo{author}{\bibfnamefont{H.~J.} \bibnamefont{Choi}},
  \bibinfo{author}{\bibfnamefont{D.}~\bibnamefont{Roundy}},
  \bibinfo{author}{\bibfnamefont{H.}~\bibnamefont{Sun}},
  \bibinfo{author}{\bibfnamefont{M.~L.} \bibnamefont{Cohen}}, \bibnamefont{and}
  \bibinfo{author}{\bibfnamefont{S.~G.} \bibnamefont{Louie}},
  \bibinfo{journal}{Nature} \textbf{\bibinfo{volume}{418}},
  \bibinfo{pages}{758} (\bibinfo{year}{2002}).

\bibitem[{\citenamefont{Zhang et~al.}(2005)\citenamefont{Zhang, Louie, and
  Cohen}}]{Zhang2005}
\bibinfo{author}{\bibfnamefont{P.}~\bibnamefont{Zhang}},
  \bibinfo{author}{\bibfnamefont{S.~G.} \bibnamefont{Louie}}, \bibnamefont{and}
  \bibinfo{author}{\bibfnamefont{M.~L.} \bibnamefont{Cohen}},
  \bibinfo{journal}{Phys. Rev. Lett.} \textbf{\bibinfo{volume}{94}},
  \bibinfo{pages}{225502} (\bibinfo{year}{2005}).

\bibitem[{\citenamefont{Crangle and Goodman}(1971)}]{crangle1971}
\bibinfo{author}{\bibfnamefont{J.}~\bibnamefont{Crangle}} \bibnamefont{and}
  \bibinfo{author}{\bibfnamefont{G.~M.} \bibnamefont{Goodman}},
  \bibinfo{journal}{Proc. R. Soc. London, Ser. A}
  \textbf{\bibinfo{volume}{321}}, \bibinfo{pages}{477} (\bibinfo{year}{1971}).

\bibitem[{\citenamefont{Tran et~al.}(2012)\citenamefont{Tran, Koller, and
  Blaha}}]{tran12}
\bibinfo{author}{\bibfnamefont{F.}~\bibnamefont{Tran}},
  \bibinfo{author}{\bibfnamefont{D.}~\bibnamefont{Koller}}, \bibnamefont{and}
  \bibinfo{author}{\bibfnamefont{P.}~\bibnamefont{Blaha}},
  \bibinfo{journal}{Phys. Rev. B} \textbf{\bibinfo{volume}{86}},
  \bibinfo{pages}{134406} (\bibinfo{year}{2012}).

\bibitem[{\citenamefont{Jang and Deok~Yu}(2012)}]{jang12}
\bibinfo{author}{\bibfnamefont{Y.-R.} \bibnamefont{Jang}} \bibnamefont{and}
  \bibinfo{author}{\bibfnamefont{B.}~\bibnamefont{Deok~Yu}},
  \bibinfo{journal}{J. Phys. Soc. Jpn.} \textbf{\bibinfo{volume}{81}},
  \bibinfo{pages}{114715} (\bibinfo{year}{2012}).

\end{thebibliography}

\end{document}